\documentstyle[twocolumn,prb,aps,epsfig,floats]{revtex}
\begin{document}
\hyphenation{presented}
\twocolumn[\hsize\textwidth\columnwidth\hsize\csname @twocolumnfalse\endcsname

\title{Hall cross size scaling and its application\\
to measurements on nanometer-size iron particle arrays}
\author{S. Wirth and S. von Moln\'ar}
\address{MARTECH, Florida State University, Tallahassee, FL 32306-4351}
\date{\today}
\maketitle

\begin{abstract}
Hall crosses were used to measure the magnetic properties of arrays of 
ferromagnetic, nanometer-scale iron particles. The arrays typically consist of
several hundred particles of 9 -- 20 nm in diameter. It is shown that the 
sensitivity of the measurements can be improved by matching the areas of the 
Hall cross and the array grown onto it by at least an order of magnitude.
We predict that single particles of diameter as small as 10 nm can be 
measured if grown onto a Hall cross of appropriate size.
\end{abstract}
]

\narrowtext
There continues to be great interest in small magnetic particles and 
well-arranged particle arrays. This research is driven by the demand for an 
understanding of the physics and potential application of such particle arrays
as advanced magnetic storage media. The challenge, however, is not only in the
fabrication of such particles but also in the measurement of magnetic 
properties of small volumes. One obvious choice for the latter is the use of 
dc micro-SQUIDs with which excellent experiments have been 
performed.\cite{wer3,wer4} These measurements, however, are ordinarily
restricted to low temperatures. MFM investigations can provide spatially
resolved information on the magnetic state of the particles and are of special
merit if conducted in applied fields. They suffer, however, in that
quantitative results are difficult to extract. Hall magnetometry has the 
advantages of being versatile\cite{lot,geim,rou,kent1} and comparatively simple
in set-up. Furthermore, temperature and magnetic field strength are not 
restricted. Applications range from flux characterization in superconductors 
to scanning Hall probes. The sensitivity and the accessible temperature range 
depend on the material used for fashioning the device. 
 
Hall gradiometry has been used to measure the magnetization of small particle 
arrays.\cite{kent2,wi1,wi2} Here, nanometer-scale particles of regular shape 
and arrangement were grown onto III-V semiconductor Hall crosses. This allowed
magnetization measurements over a broad temperature range (up to 100 K). 
Moreover, enhanced interactions between the particles have been 
studied.\cite{wi3} Thus, important issues in data storage applications 
(superparamagnetic limit, interactions) can be addressed.

In this letter, we show that the sensitivity of Hall measurements can be
significantly increased by matching the sizes of the active area of the Hall 
cross and of the particle array. Hall voltages calculated\cite{wi1} from the
magnetic stray field of the particles are compared to measured values. We 
predict that single nanometer-size particles can be measured by appropriately 
small Hall crosses. The results can be applied to any magnetic object to be 
measured by Hall magnetometry.

Iron particle arrays were grown by a combination of chemical vapor deposition
(CVD) and scanning tunneling microscopy (STM).\cite{wi1,wi2,kent3} This method
has successfully been used to fabricate particles from 9 to 20 nm in diameter,
50 - 250 nm in height and with interparticle distances down to 80 nm onto gold
and permalloy. During the growth process, vaporous iron pentacarbonyl is 
introduced into the STM chamber and decomposed within the electric field of 
the biased tip. For negative tip bias voltage, the iron deposit grows on the 
sample surface. At the same time, the tip is retracted, keeping the distance 
between the deposit's top and the tip constant by maintaining a constant 
tunneling current of 50 pA. When the deposit has grown to the desired height 
(measured via the tip retraction) the tip is retracted completely and moved 
to the next location on the sample surface where the process is repeated to  
\begin{figure}[!b]
\centering
\epsfig{figure=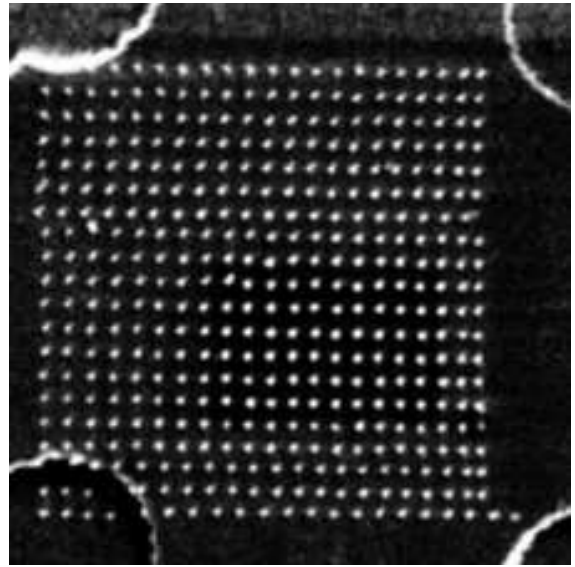,width=7.5cm}
\vspace*{0.3cm}
\caption{SEM picture of an array of 420 particles grown onto a Hall cross. The
etched Hall cross of about $3.2 \times 2.8\: \mu$m$^2$ is clearly visible. The
image shows an area of $4.5 \times 4.5\: \mu$m$^2$.}
\label{sempic}
\end{figure}
form the particle array. An advantage of this fabrication procedure is that  
the particles' height and location (with respect to each other and to features 
on the sample surface) can easily be controlled by steering the STM-tip. This 
feature becomes even more important as the size of the Hall crosses is
decreased. It should be noted that the magnetic cores of the particles 
(consisting of bcc iron as revealed by TEM\cite{kent3}) are surrounded by a
carbon coating which reduces oxidation and aging of the samples under air.

For the magnetic measurements, III-V semiconductor Hall crosses were prepared
by photolithography and wet chemical etching of the substrate 
(GaAs-Ga$_{0.7}$Al$_{0.3}$As two-dimensional electron system (2DES), 
$n_{2D}$ = 1.2 $\times\: 10^{11}$cm$^{-2}$, $\mu$(30$\:$K) = 4.5 $\times \:
10^5$ cm$^2$/Vs). A 40 nm thin gate was deposited onto the Hall bars, on top of 
which the arrays were directly grown. The SEM image of a typical array is
presented in Fig.~\ref{sempic}. 

The measured Hall voltage originates from the magnetic stray field of the 
particles. A two-step procedure was developed\cite{wi1} to analyze the Hall 
voltage: In a first step, the stray field (i.e., its $z$-component perpendicular 
\begin{figure}[t]
\centering
\epsfig{figure=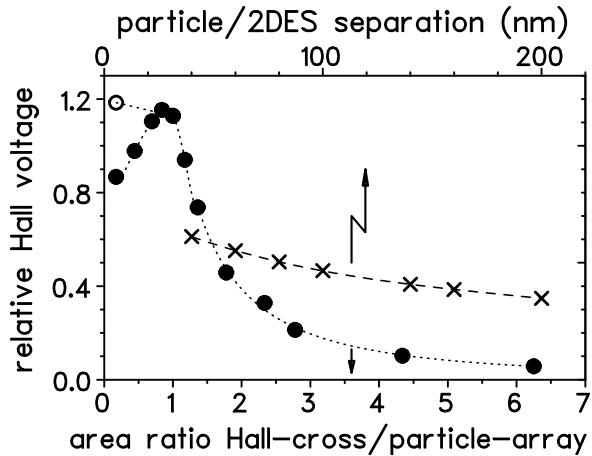,width=8cm}
\vspace*{0.3cm}
\caption{Calculated relative Hall voltages of a typical test array. Hall 
crosses of different sizes were assumed (circles, lower axis). Array and
Hall cross were either assumed to be aligned with their centers (closed 
circles) or with one of their corners (open circle). In addition, the relative
Hall voltage for different separations between particles and the 2DES is shown 
(crosses, upper axis).}
\label{volt}
\end{figure}
to the plane of the 2DES) of each particle at the depth at which the 2DES 
located is calculated analytically. The contributions of each particle are then 
summed up to get the local $z$-component of the stray field emanating from the 
whole array. In a second step, the Hall voltage produced by this magnetic stray
field within the active area of the Hall cross is calculated. The parameters 
needed are either known from the fabrication process of the 2DES or can easily 
be measured. With all other parameters known, the procedure can be used to
estimate the mean diameter of the particle iron core from the measured Hall
voltage.

In order to optimize the sensitivity of the Hall device the efficiency of 
the conversion of the particles' stray field into Hall voltage, i.e. the second
step, has to be evaluated. Properties most easily controlled in the 
fabrication process are the relative size and location of the array with 
respect to the Hall cross (we do not intend 
to discuss the properties of different 2DES materials in this letter). In 
Fig.~\ref{volt} (circles) the calculated, relative Hall voltage produced by a
typical test array is presented if the array were grown onto Hall crosses of 
different sizes (all other parameters remained unchanged including the drive 
current). Obviously, the array's stray field is most effectively converted 
into Hall voltage if the Hall cross size does not exceed the array size. In 
this case, all electrons in the active area of the Hall cross are influenced 
by the stray field and the stray field influences the potential over the
complete width of the voltage leg. Compared to arrays fabricated 
earlier\cite{kent2,wi2} one should be able to increase the sensitivity by an 
order of magnitude just by matching the sizes of array and Hall cross. The
calculations were performed assuming aligned centers of array and Hall cross 
(filled circles). If the Hall cross is smaller than the array, only that 
corresponding portion of the array causes a Hall voltage. The total stray
field at the center of an array is, however, slightly smaller than at an 
edge or even at a corner of the array. This effect which has been noted
before\cite{gid} can be explained by the fact that the closest (and 
therefore most effective) fluxlines from center particles form closed lines
within the active area of the Hall cross and do not contribute to the Hall 
voltage. Hence, the Hall response is reduced at the array's center. In 
contrast, for a small Hall cross located with its edge underneath the 
corresponding edge of the array, the Hall response is slightly increased.

Another parameter that could somewhat be influenced is the separation in 
$z$-direction between the 2DES layer and the particles (e.g. via the gate 
thickness). This, however, is expected to have only a minor influence on the 
Hall response (Fig.~\ref{volt}, crosses and upper scale).

To make use of the predicted increase in sensitivity we prepared Hall crosses
of approximately 3 $\times$ 3 $\mu$m$^2$ in size (cf. Fig.~\ref{sempic}). As 
described before, arrays of several hundred iron particles with interparticle 
distances of 150 nm were then grown onto these Hall crosses. The dimensions 
\begin{figure}[t]
\centering
\epsfig{figure=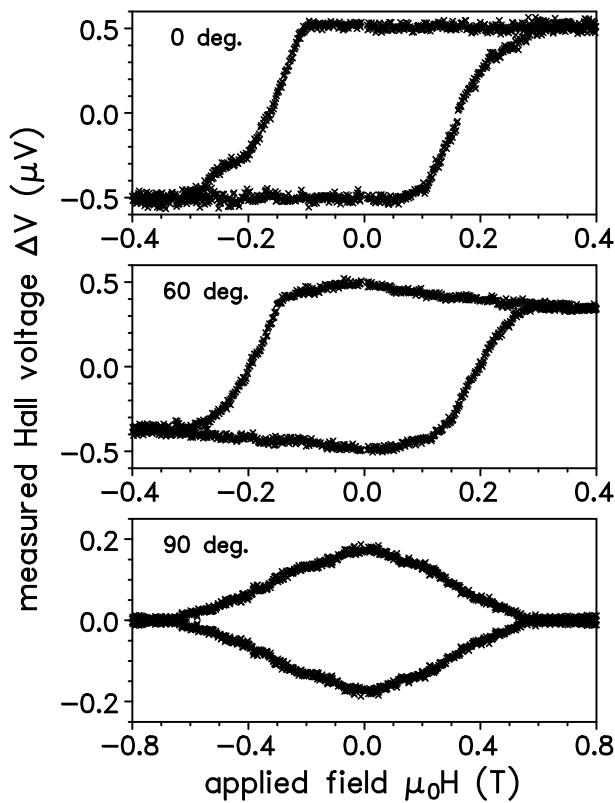,width=8.2cm}
\vspace*{0.3cm}
\caption{Hall voltages of an array of 462 particles with a 17 nm diameter 
measured at 30 K. The field was applied at different angles (shown are 
0$^{\circ}$, 60$^{\circ}$, 90$^{\circ}$) with respect to the particles' long
axis, i.e. their easy magnetization direction.} 
\label{meas}
\end{figure}
were chosen to match the size of the Hall crosses to be grown onto and to 
approximate arrays fabricated earlier.\cite{wi1,wi2} This permits direct 
comparison of the measured Hall voltages. Magnetic measurements were performed
from 10 -- 100 K and for different angles of the applied field. Due to their 
elongated shape (aspect ratio of approx. 5:1) the particles possess a large
shape anisotropy making their long axis an easy magnetization direction (EMD) 
along $z$. A mean particle switching field of $H_{sw}\,=\:$160 mT for fields 
applied parallel to the EMD (Fig.~\ref{meas}, top panel) was observed. Apart
from the typical switching field distribution (about 30 mT) there was a small
portion of particles with a distinguishable higher $H_{sw}$ of about 230 mT 
as indicated by ``bumps'' in the corresponding parts of the magnetization 
curves. Such a twofold switching field distribution may be caused by a 
distribution of the magnetic core diameter of the particles. It seems more
likely, however, that the majority of the particles consisted of more than one
grain. Therefore, they have a smaller switching field than single crystalline 
particles. Such structural differences would naturally account for 
distinguishable $H_{sw}$-values.
 
For fields perpendicular to the EMD, the magnetization behavior was controlled
by reversible rotation (Fig.~\ref{meas}, lower panel). For increasing fields 
strength, the particle magnetization rotated toward the field direction,
i.e. toward an orientation perpendicular to the $z$-direction. Thus, a 
decreasing Hall voltage is measured. For intermediate angle (shown in the 
center panel of Fig.~\ref{meas} is the curve measured at 60$^{\circ}$) both 
reversible rotation and switching contribute to the overall magnetization 
behavior: the former ends to a small decrease of the $z$-component of the
magnetization (visible at fields below about 130 mT) whereas the latter 
dominates around 200 mT. An estimate of the mean value of the shape anisotropy
constant from the reversible part of this curve (based on the Stoner-Wohlfarth 
model) yielded $K_S \approx 0.3$ MJm$^{-3}$. This value is about 40\% of the 
number expected from the particles' shape. This again indicates that most of
the particles are polycrystalline. The peculiar shape of the curve measured 
at 90$^{\circ}$ can then be explained by a distribution of $K_S$-values with 
maximum $K_S$-values as high as 0.45 MJm$^{-3}$.

The Hall voltages measured for 0$^{\circ}$ and 60$^{\circ}$ exceeded those 
measured earlier\cite{wi1,wi2} by more than an order of magnitude (after 
adjusting for changed experimental conditions, e.g. drive current, carrier
concentration) in good agreement with our predictions. At zero field, the 
measured Hall voltage should not depend on the orientation of the applied 
field if all magnetizations of the individual particles point in the same
direction. Obviously, this condition was not fulfilled for curves measured
for 90$^{\circ}$. In fact, for fields decreasing from a value well above the 
anisotropy field the particle magnetization could rotate toward either 
direction along the EMD. From the 0$^{\circ}$ and 60$^{\circ}$-curves a mean
particle core diameter of 17 nm was estimated (particles were grown 80 nm in 
height). 

We emphasize that the resulting Hall voltage does not substantially depend on 
the {\em absolute} size of the array or the Hall cross--as long as they match.
In the present experiment the noise of the Hall voltage was measured to be 
0.04 $\mu$V$/\sqrt{Hz}$ at 30 K and zero field and increased to about 0.07 
$\mu$V$/\sqrt{Hz}$ at 100 K and 1.0 T. We predict a Hall voltage of 
$\approx\, 0.24\:\mu$V for a single particle of 10 nm diameter grown onto a 
$400 \times 400$ nm$^2$ Hall cross (experimental conditions as for the 
measurements in Fig.~\ref{meas}, no depletion effects of the 2DES were taken
into consideration). This voltage would exceed the highest noise level by a
factor of 5. Hence, we expect to be able to measure any number of particles, 
from a single particle to a few particles up to arrays of several hundred
particles, by growing them on Hall crosses of appropriate size. Here, STM 
assisted growth appears to be the perfect tool. 

The authors wish to thank A. C. Gossard for providing the 2DES material and
A. Anane for helpful discussions.


\vspace*{-0.4cm}
\begin{references}
\bibitem{wer3} W. Wernsdorfer, B. Doudin, D. Mailly, K. Hasselbach, A. Benoit,
J. Meier, J.-P. Ansermet and B. Barbara, Phys. Rev. Lett. {\bf 77}, 1873 
(1996).

\bibitem{wer4} W. Wernsdorfer {\it et al.}, 
Phys. Rev. Lett. {\bf 78}, 1791 (1997). 

\bibitem{lot} D. Lottis, F. Petroff, A. Fert and M. Konczykowski, J. Magn.
Magn. Mater. {\bf 104-7}, 1811 (1992).

\bibitem{geim} A. K. Geim {\it et al.}, Nature, {\bf 390}, 259 (1997).

\bibitem{rou} F. G. Monzon, D. S. Patterson and M. L. Roukes, J. Magn. Magn.
Mater. {\bf 195}, 19 (1999).

\bibitem{kent1} A. D. Kent {\it et al.}, 
submitted to Europhys. Lett.

\bibitem{kent2} A. D. Kent, S. von Moln\'ar, S. Gider and D. D. Awschalom, 
J. Appl. Phys. {\bf 76}, 6656 (1994).

\bibitem{wi1} S. Wirth, M. Field, D. D. Awschalom and S. von Moln\'ar,
Phys. Rev. B {\bf 57}, R14028 (1998).

\bibitem{wi2} S. Wirth, M. Field, D. D. Awschalom and S. von Moln\'ar, J. Appl.
Phys. {\bf 85}, 5249 (1999).

\bibitem{wi3} S. Wirth and S. von Moln\'ar, submitted to J. Appl. Phys.

\bibitem{kent3} A. D. Kent, T. M. Shaw, S. von Moln\'ar and D. D. Awschalom, 
Science {\bf 262}, 1249 (1993).

\bibitem{gid} S. Gider {\it et al.}, 
Appl. Phys. Lett. {\bf 69}, 3269 (1996). 

\end{references}
\end{document}